\documentclass[11pt]{article}
\usepackage{moriond-photo,epsfig,amsmath}
\usepackage{wrapfig}
\usepackage{url}

\newcommand{\ie}{i.e.\ }
\newcommand{\as}{\alpha_s}
\newcommand{\GeV}{\,\mathrm{GeV}}
\newcommand{\order}[1]{{\cal O}\left(#1\right)}

\begin{document}
\vspace*{4cm}
\title{A Practical Seedless Infrared Safe Cone
  Algorithm\,\footnote{Talk presented at the XLII Rencontres de
    Moriond, QCD and Hadronic Interactions, La Thuile, March 2007.}}

\author{ Gavin P.~Salam }

\address{LPTHE, 
  Universit\'e Pierre et Marie Curie -- Paris 6,\\
  Universit\'e Denis Diderot -- Paris 7,
  CNRS UMR 7589, 75252 Paris cedex 05, France.}

\maketitle\abstracts{ This writeup highlights the infrared unsafety of
  the `midpoint' cone jet-algorithm and provides a brief overview of why this
  is a serious issue. It then shows how one can build a safe
  (seedless) cone algorithm and discusses the potential impact on
  measurements.  }

Two broad classes of jet algorithm are in widespread use at modern
colliders: sequential recombination type algorithms, such
as $k_t$\,\cite{Kt} and Cambridge/Aachen,\cite{Cam} and cone-type
ones.\cite{Blazey} The former take a 
bottom-up approach to the problem of defining jets, repeatedly
combining particles that are closest in some distance measure. They
work because the proximity measures used are closely related with QCD
divergences for particle production, and they are much appreciated in
the $e^+e^-$ and $ep$ communities, both for their simplicity and their modest
hadronisation corrections. Cone type algorithms take a top-down
approach, finding coarse regions of energy flow (cones) and
identifying them as jets.  They work because QCD only modifies the
energy flow on small scales, and so far they have been the preferred
type of algorithm in the $pp$ community, because of the greater
geometrical regularity of the resulting jets and their sometimes lower
sensitivity to certain components of the non-perturbative underlying
event and pileup.

Cone algorithms have been in use since the early 1980's,\cite{UACone}
and in the early 1990's awareness developed\,\cite{Snowmass} of the
importance for cone-algorithm formulations to satisfy a certain basic
set of requirements: they must be fully defined, practical in both
experimental and theoretical contexts, and cross sections must be
finite at any order of perturbation theory, \ie the algorithm must be
infrared and collinear (IRC) safe.

Modern cone algorithms involve two main steps: a procedure to find
`stable cones' (a cone pointing in the same direction as the momentum
of its contents) and a `split--merge' procedure to convert those cones
into jets, resolving the problem of cones that have particles in
common (\ie that `overlap'). 

The most delicate issue with cone algorithms has been that of finding
the stable cones. A standard procedure had long been to use all
particles (possibly above a seed threshold) as directions of trial
cones, then for each trial cone to use the momentum of its contents as
a new trial direction, iterating until stable directions are obtained.
The drawback of iterative procedures is that new stable cones (and
jets) may be found if an extra starting point is added. This was known
to happen with the addition of soft particles in straightforward
iterative stable-cone searches,\cite{midpoint} but it had been thought
that a trick of adding extra starting points, at the midpoints between
the cones already found, would lead to a final set of stable cones that
was insensitive to the addition of extra seeds. Accordingly a
recommendation was made\,\cite{Blazey} for the Tevatron experiments to
use such a `midpoint' iterative cone algorithm.

\begin{wrapfigure}{R}{0pt}
  \includegraphics[width=0.55\textwidth]{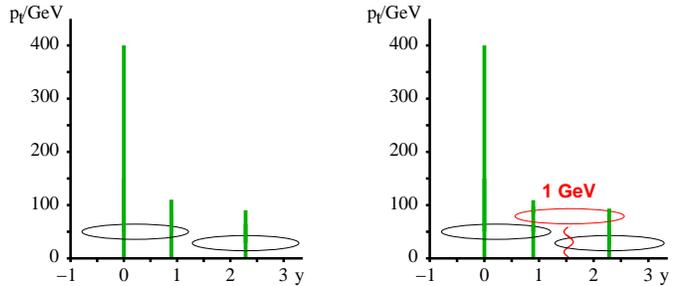}
  \caption{A configuration (left) in which the midpoint algorithm
    ($R=1$) gives different jets if a soft particle is added (right).}
  \label{fig:problem}
\end{wrapfigure}
It turns out that while the midpoint fix resolves the infrared
problems for events with two neighbouring hard particles, those
problems reappear for three or more neighbouring hard particles.\cite{TeV4LHC,SISCone}
This is illustrated in fig.\ref{fig:problem} where in the left-hand
particle configuration two stable cones (and so two jets) are found
with the midpoint cone algorithm. If a soft, $\sim 1\GeV$, particle is
now added (right), it provides an extra seed leading to a third
(overlapping) stable cone being found, and all the cones are then
merged into a single jet (for $f=0.5$).

\begin{table}[b]
  \centering
      \begin{tabular}{|l|c|c|}\hline
        Observable                                & 1st miss cones at & Last meaningful order \\ \hline
        Inclusive jet cross section               &  NNLO             & NLO \\ 
        $W/Z/H$ + 1 jet cross section             &  NNLO             & NLO \\
        $3$ jet       cross section               &   NLO             & LO  \\
        $W/Z/H$ + 2 jet cross section             &   NLO             & LO  \\
        jet masses in $3$~jets, $W/Z/H + 2$~jets  &    LO             & none \\\hline
      \end{tabular}
      \caption{Summary of the order ($\as^4$ or $\as^3 \alpha_{EW}$)
        at which stable cones are missed for 
        various observables with a midpoint algorithm, and the corresponding
        last order that can be meaningfully calculated. 
        (Legacy iterative cone algorithms, without midpoint seeds, such as
        JetClu, fail one order earlier).
      }
  \label{tab:failure-cases}
\end{table}

The sensitivity to the set of seeds means that the midpoint cone
algorithm is either infrared unsafe (without a seed threshold) or
collinear unsafe (with a seed threshold). This is a serious issue, for
many reasons: 1) it defeats the purpose of using a jet algorithm in
the first place: a jet algorithm is supposed to provide a
correspondence between the complex hadron level and a simple
few-parton picture of an event --- this correspondence is meaningless
if a random $1\GeV$ non-perturbative particle changes the
multi-hundred $\GeV$ jets. 2) IRC unsafety invalidates the theorems
that ensure the finiteness of perturbative QCD calculations, because
the jets found in (divergent, supposedly cancelling) real and virtual
diagrams differ.  3) Pragmatically it limits the accuracy with
which one can meaningfully predict many observables, as summarised in
table~\ref{tab:failure-cases}, and already programs such as
NLOJET\,\cite{NLOJet} or MCFM\,\cite{MCFM} allow one to go beyond this
order when using a safe jet algorithm. Therefore the use of a midpoint
algorithm squanders the potential for accurate predictions that stems
from many years of hard theoretical calculations, and forever limits the
usefulness of data measured with it.

A solution to the cone algorithm's problems is to carry out an
exhaustive (`seedless') search for all stable cones. Since additional
soft particles do not change the stability of cones, if one has already
found all stable cones adding a soft particle cannot lead to extra stable
cones being found, and so the IRC safety problem is eliminated. A
seedless algorithm had been proposed\,\cite{KOS,Blazey} for perturbative
calculations, but since it took time $\sim N 2^N$ to find jets among $N$
particles ($10^{17}$ years for $N=100$), it was unthinkable to use it
at hadron or detector level.

\begin{figure}
  \centering
  \includegraphics[width=\textwidth]{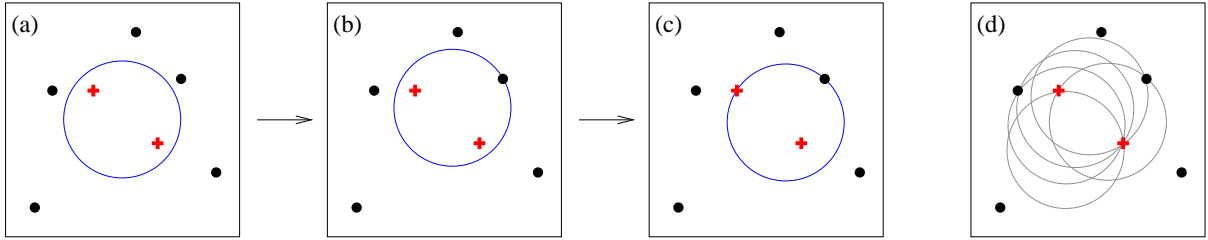}
  \caption{(a) Some initial circular enclosure; (b) moving the circle
    in a random direction until some enclosed or external point
    touches the edge of the circle; (c) pivoting the circle around the
    edge point until a second point touches the edge; (d) all circles
    defined by pairs of edge points leading to the same circular
    enclosure.}
  \label{fig:2dcircle}
\end{figure}

\begin{wrapfigure}{R}{0pt}
  \includegraphics[width=0.5\textwidth,angle=270]{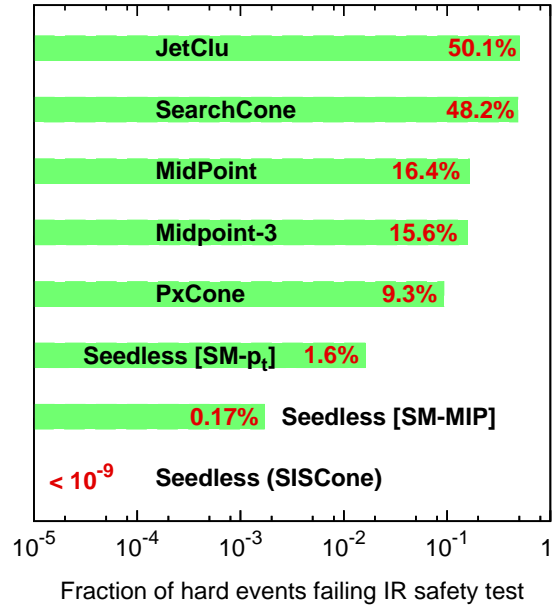}
  \caption{Failure rates for the IR safety tests with various
    algorithms, including a midpoint variant with $3$-way midpoints
    and some seedless algorithms with commonly used, but improper,
    split--merge procedures. }
  \label{fig:IR_failures}
\end{wrapfigure}
Recently it was observed\,\cite{FastJet} that it can be advantageous
to relate sequential-recombination jet algorithms to problems in
computational geometry. It turns out that this is true also of cone
algorithms, for which the exhaustive stable cone search reduces to a
2D `all distinct circular enclosures' problem. While apparently not
having been considered by the computational geometry community, this
problem is easily solved, essentially by considering all circles
having a pair of particles on their circumference, cf.\
fig.~\ref{fig:2dcircle}. With the aid of further standard
computational techniques one obtains\,\cite{SISCone} a seedless
algorithm that takes $\order{N^2 \ln N}$ time. Not only does this
provide a practically usable IR safe cone algorithm, but it even
scales better at large $N$ than midpoint algorithms ($N^3$) and is of
similar speed to them for the values of $N \sim 500-1000$ that will be
found at low-luminosity LHC. 

Given the cone algorithm's chequered history with IRC safety, it is
important to establish, as far as possible, that there are no further
unpleasant surprises waiting to be discovered in a few years' time.
This has been done in two ways: with a detailed analytical proof, and
via Monte Carlo tests in which one finds jets in a `hard event' (with up
to 10 hard particles), repeatedly adds infinitely soft particles and
verifies that the jets found are the same. If they are not, then the
algorithm is IR unsafe. The failure rates on this test are shown for a
variety of cone algorithms in fig.~\ref{fig:IR_failures}. Among the
discoveries made in these tests, was that the split--merge procedure
also had the potential to create IR safety problems. The final version
of the seedless algorithm, named SISCone, has passed several billion
hard event tests without failure. The code for the algorithm is
available publicly\,\cite{code} both in standalone form and as a
FastJet\,\cite{FastJet} plugin.

\begin{wrapfigure}{R}{0.5\textwidth}
  \includegraphics[width=0.5\textwidth]{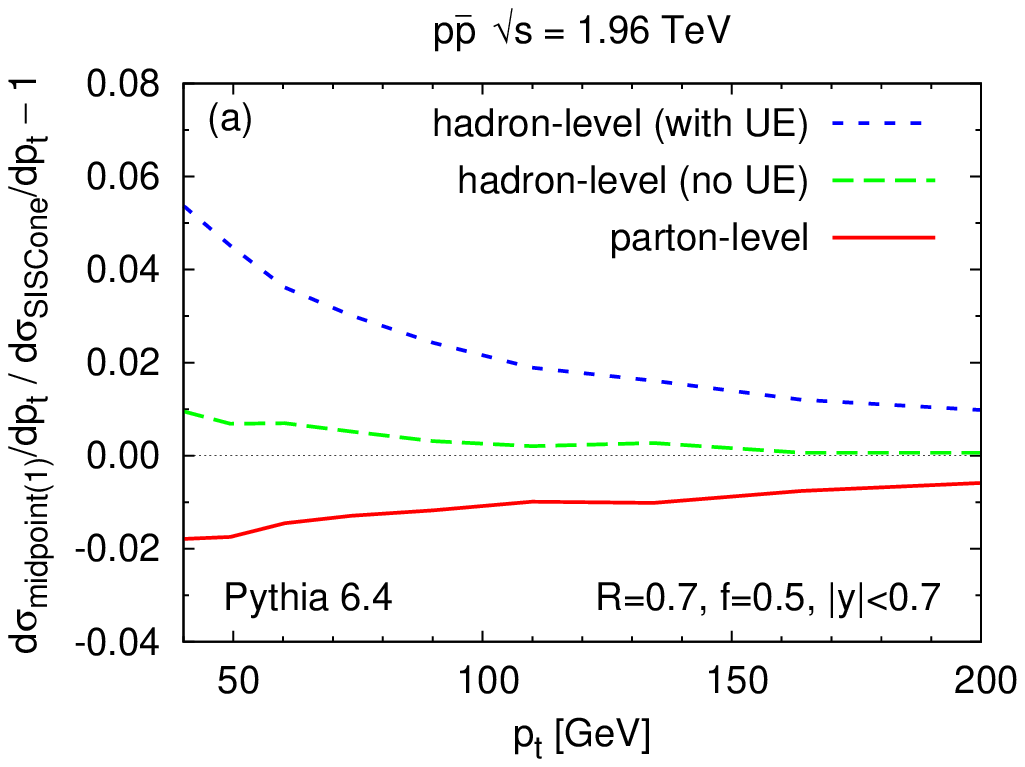}\\
  \mbox{ }\hfill \includegraphics[width=0.47\textwidth]{nlojet_mass2_less2R.ps}
  \caption{Top: relative difference between the midpoint and SISCone
    inclusive jet $p_t$ spectra at the Tevatron. Bottom: relative
    difference between the midpoint and SISCone jet mass spectra, in
    $3$-jet events for which the second and third hardest jets are
    in a common neighbourhood.\vspace{-5em}}
  \label{fig:physics}
\end{wrapfigure}

The physics impact of switching from the midpoint to SISCone depends
on the observable and is illustrated in fig.~\ref{fig:physics} for two
cases. For inclusive quantities, like the inclusive jet spectrum
(upper panel), one sees effects of the order of a couple of percent,
as is to be expected since stable cones are only missed at NNLO
onwards. One notes nevertheless that differences of up to $5\%$ arise
when including underlying event effects, and this is related to
SISCone's substantially lower sensitivity to diffuse `noise' in an
event.

For more exclusive quantities the differences between midpoint and
SISCone are more significant. For jet-mass spectra in three-jet
events (lower panel of fig.~\ref{fig:physics}), the
difference starts are LO, and this can translate to $40\%$ effects in
partonic predictions (which essentially corresponds to an unavoidable
$40\%$ non-perturbative ambiguity for the midpoint algorithm).

To conclude, while both sequential recombination and cone-type jet
algorithms have their place at hadron colliders, it is essential that
they be practical and safely defined. The widespread `midpoint' cone
algorithm is not infrared safe, and therefore there are strong reasons
for discontinuing its use in favour of a seedless cone algorithm such
as SISCone, which is both infrared safe and practical at parton,
hadron and detector levels.

\section*{Acknowledgements}
This work was carried out in collaboration with Gregory Soyez and
supported in part by grant ANR-05-JCJC-0046-01 from the French Agence
Nationale de la Recherche.


\end{document}